\documentclass[twocolumn,showpacs,preprintnumbers,amsmath,amssymb,superscriptaddress]{revtex4}

\usepackage{amsmath,amsfonts,amssymb}
\usepackage[english]{babel} 
\usepackage[latin1]{inputenc} 
\usepackage[T1]{fontenc}
\usepackage{color}
\usepackage{float}
\usepackage{verbatim}
\usepackage{graphicx}
\usepackage{bm}
\usepackage{mathtools}

\def \equi#1{\mathrel{\mathop{\kern 0pt\sim}\limits_{#1}}}

\newcommand{\diff}[1]{\mathrm{d}#1 \;}
\newcommand{\e}[1]{\; \mathrm{e}^{#1} \;}
\newcommand{\erfc}[1]{\; \mathrm{erfc}\left(#1\right) \;}

\begin{document}

\title{Convex hull of a Brownian motion in confinement}

\author{Marie Chupeau}
\affiliation{Laboratoire de Physique Th\'eorique de la Mati\`ere Condens\'ee (UMR CNRS 7600), 
Universit\'e Pierre et Marie Curie, 4 Place Jussieu, 75255 Paris Cedex France}

\author{Olivier B\'enichou}
\affiliation{Laboratoire de Physique Th\'eorique de la Mati\`ere 
Condens\'ee (UMR CNRS 7600), Universit\'e Pierre et Marie Curie, 4 
Place Jussieu, 75255 Paris Cedex France}

\author{Satya N. Majumdar}
\affiliation{CNRS, LPTMS, Univ. Paris-Sud, 91405 Orsay Cedex, France}

\begin{abstract}
We study the effect of confinement on the mean perimeter of the convex hull of a planar Brownian motion,
defined as the minimum convex polygon enclosing the trajectory.
We use a minimal model where an infinite reflecting wall confines the walk to its one side. We show that the
mean perimeter displays a surprising minimum with respect to the starting distance to the wall and
exhibits a non-analyticity for small distances. In addition, the mean span of the trajectory in a fixed direction
\mbox{$\theta \in ]0,\pi/2[$}, which can be shown to yield the mean perimeter by integration over
$\theta$, presents these same two characteristics. This is in striking contrast with the one dimensional
case, where the mean span is an increasing analytical function. The non-monotonicity in the 2D case
originates from the competition between two antagonistic effects due to the presence of the wall:
reduction of the space accessible to the Brownian motion and effective repulsion.

\end{abstract}


\maketitle

How does one characterize the territory covered by a Brownian motion in two dimensions?
This question naturally arises in ecology where the trajectory of an animal
during the foraging period is well approximated by a Brownian motion~\cite{Berg:1983,Bartumeus:2005} 
and one needs to estimate the home range of the animal, i.e., the two dimensional (2D)
space over which the animal moves around over a fixed period of time~\cite{Murphy:1992}.
The most versatile and popular method to characterize the home range consists
in drawing the convex hull, i.e., the minimum convex polygon enclosing the
trajectory of the animal~\cite{Worton:1995,Giuggioli:2011}.
The size of the home range is then naturally estimated by the mean perimeter or the
mean area of the convex hull. 

For a single planar Brownian motion of duration $t$ and diffusion constant $D$, the
mean perimeter \mbox{$\langle L(t)\rangle = \sqrt{16\, \pi\, D\, t}$} and the mean area \mbox{$\langle A(t)\rangle=
\pi\, D\, t$} were
computed exactly in the mathematics literature quite a while 
back~\cite{Takacs:1980,ElBachir:1983,Letac:1993}. 
Very recently, there has been a growing number of articles both in the physics~\cite{MajumdarPRL09,
Majumdar:2010,Reymbaut:2011,Dumonteil:2013,Lukovic:2013}
and the mathematics literature~\cite{Biane:2011,Eldan:2011,Kampf:2012,Kabluchko:2014} 
generalizing these results in various directions.
In particular, adapting Cauchy's formula~\cite{Cauchy:1832} for closed 2D convex curves 
to the case of random curves, a general method was recently proposed~\cite{MajumdarPRL09,Majumdar:2010}
to compute the
mean perimeter and the mean area of the convex hull of any arbitrary stochastic
process in 2D. In cases where the process is isotropic in 2D, the mean perimeter
and the mean area of its convex hull can be mapped onto computing the extremal statistics of the corresponding
one dimensional component process~\cite{MajumdarPRL09,Majumdar:2010}. This procedure was
then successfully used to compute exactly the mean perimeter and the mean area of     
a number of isotropic 2D stochastic processes such as $N$ independent Brownian
motions~\cite{MajumdarPRL09,Majumdar:2010}, random acceleration process~\cite{Reymbaut:2011},
branching Brownian motion with absorption with applications to epidemic spread~\cite{Dumonteil:2013}
and for anomalous diffusion processes~\cite{Lukovic:2013}.

All these results pertain to isotropic stochastic processes in the
unconfined 2D geometry. However, in many practical situations, the stochastic
process takes place in a confined geometry. For example, the home range of animals living
in an habitat can get limited by the presence and development of urban areas nearby that
may impede the free movement of animals.
How does the confinement of the allowed space affect the size of the home range? Beyond this ecological motivation, determining the mean perimeter of the convex hull in confinement is a key question in the context of Brownian motion theory.

\begin{figure}[h]
\begin{minipage}{0.4\linewidth}
\centering
\includegraphics[width=100pt]{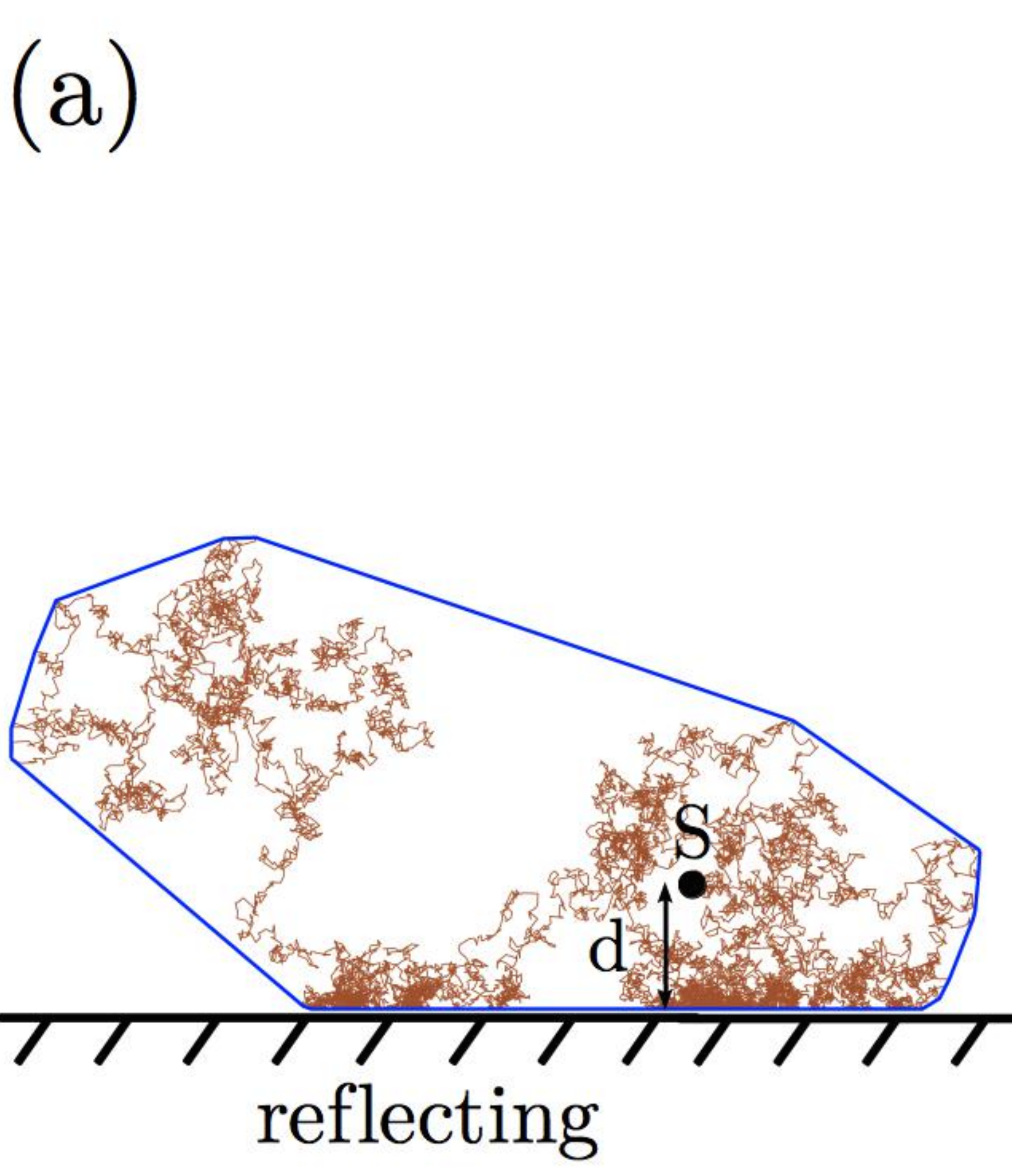}
\end{minipage} \hfill
\begin{minipage}{0.55\linewidth}
\centering
\includegraphics[width=125pt]{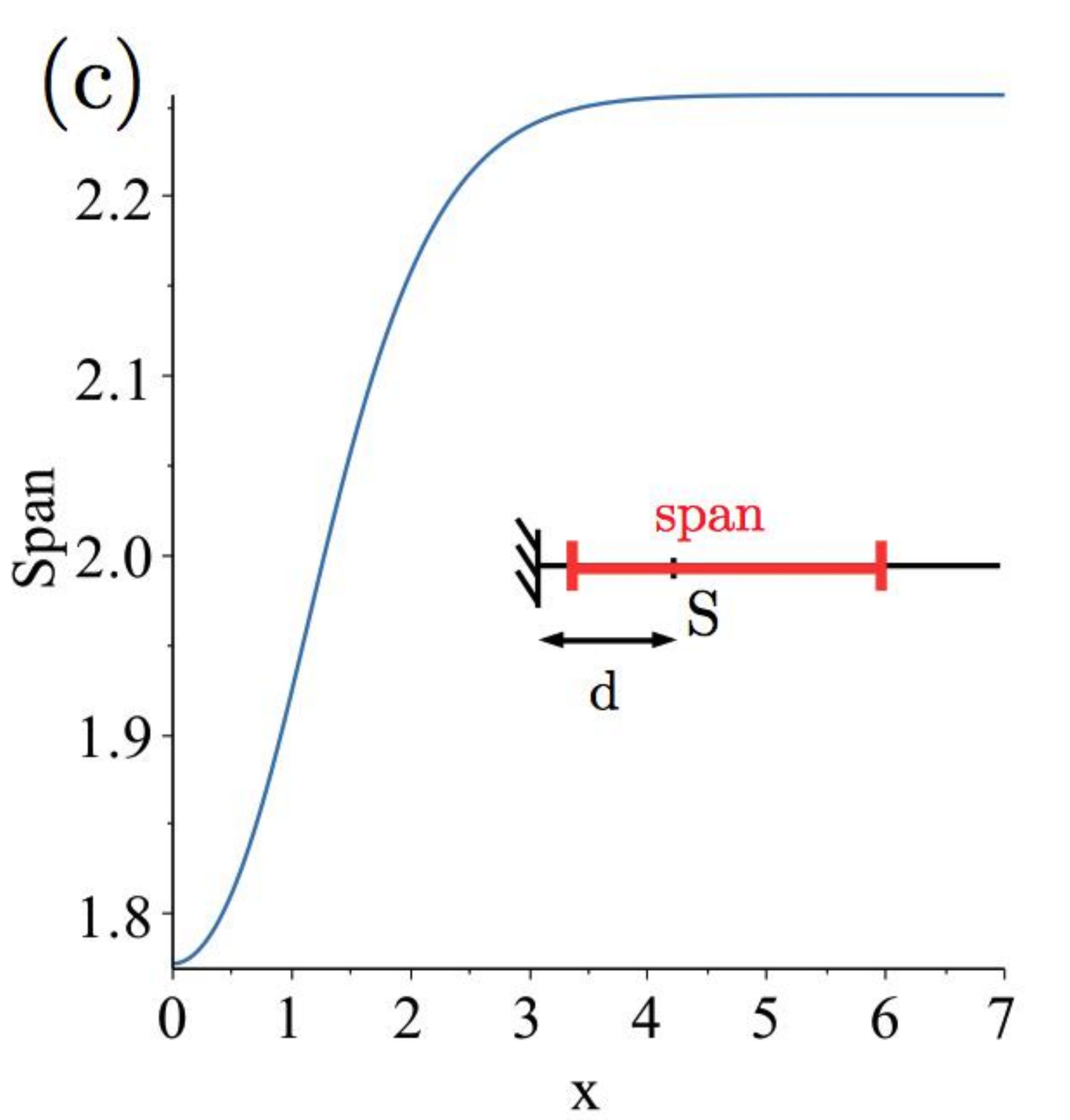}
\end{minipage}
\begin{minipage}{0.95\linewidth}
\centering
\includegraphics[width=170pt]{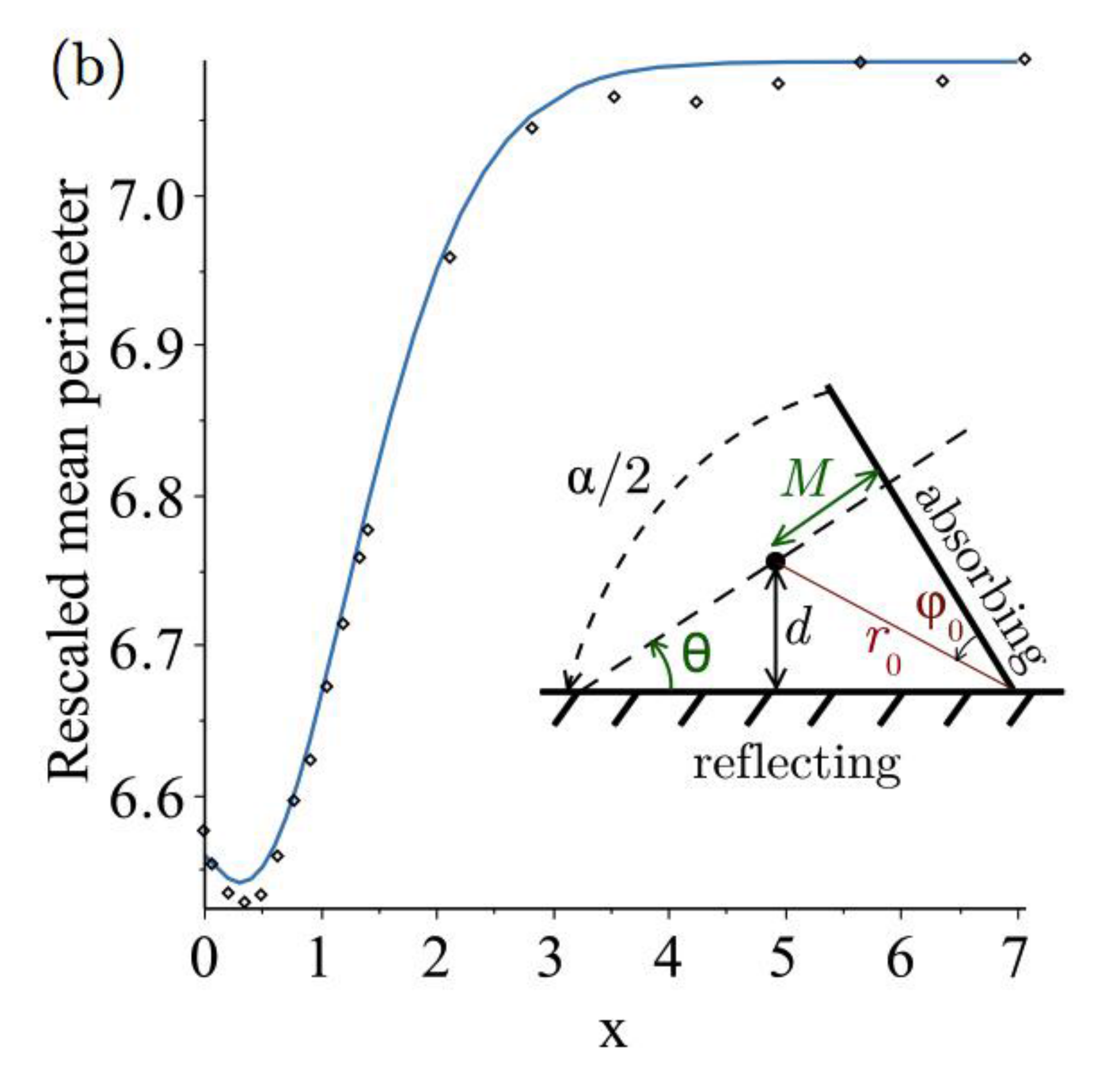}
\end{minipage}\hfill
\caption{What 
is the influence of a confinement on the convex hull of a 2D Brownian motion? Considering the minimal 
model of a reflecting confining infinite wall (see (a)), we show that the mean perimeter of the convex 
hull turns out to be a non-monotonic function of the initial distance $d$ to the wall (see (b)). This is 
in striking contrast with the monotonic behavior of the mean span, which is the 1D analogue of the mean 
perimeter of the convex hull (see (c)). \textbf{(a)} Convex hull of a Brownian trajectory of starting 
point O and diffusion coefficient $D$. \textbf{(b)} Mean rescaled perimeter $\left\langle 
L_{\mathrm{2D}}^{(d)}(t)/\sqrt{Dt} \right\rangle$ of the convex hull as a function of the rescaled distance to the 
wall $x=d/\sqrt{Dt}$: analytical (solid line) and simulation (dots) results (see \cite{SI} for details). The inset presents the 
different geometrical parameters involved in the calculation. \textbf{(c)} Mean rescaled span $\left\langle 
L_{\mathrm{1D}}^{(d)}(t)/\sqrt{Dt} \right\rangle$ of a 
(semi) confined 1D Brownian motion as a function of the rescaled distance $x=d/\sqrt{Dt}$ to the reflecting point.}
\label{recap}
\end{figure}

In this Letter, we address this important issue in a simple minimal model
that allows an exact solution. We consider a single planar Brownian motion
in the presence of a reflecting infinite wall that confines the Brownian motion
in the positive half-space (see Fig.\ref{recap}(a)). This can simply model the habitat of an animal
bordering, on one side, a highway or a river that the animal
can not cross. We show that the presence of the
wall breaks the isotropy of the 2D space in a way that, even in this simple setting, induces a nontrivial effect on the
convex hull of the Brownian motion. 
We compute analytically the mean perimeter $\langle L^{(d)}_{2 \mathrm{D}}(t)\rangle$ of the convex hull
of the Brownian motion of duration $t$, starting at an initial distance $d$ 
from the wall. We show that it exhibits a scaling form, at all times $t$,
\begin{equation}
\langle L^{(d)}_{2 \mathrm{D}}(t)\rangle=\sqrt{D\,t}\,\, {\tilde L}_{2 \mathrm{D}}\left(\frac{d}{\sqrt{D\,t}}\right)
\label{scalingform1}
\end{equation}
where ${\tilde L}_{2 \mathrm{D}}(x)$ is a nontrivial scaling function that we compute explicitly and
is plotted in Fig.\ref{recap}(b).
We find a {\em surprising}, and rather striking, non-monotonic behavior
of the scaling function ${\tilde L}_{2 \mathrm{D}}(x)$, revealing a minimum
at a certain characteristic scaled distance (see Fig.\ref{recap}(b)). Moreover,
the scaling function also exhibits an unexpected singularity as $x\to 0$, ${\tilde L}_{\rm 2D}(x)\sim -
a\, [x/\ln (1/x)]^{3/2}$ with the prefactor $a= 8\sqrt{2\,\pi^3}/[3\Gamma(3/4)]$.
We demonstrate that both the non-monotonicity and the small $x$ singularity
of the scaling function 
are purely 2D effects and are, in fact, absent in 1D (as shown in Fig.\ref{recap}(c)).  

{\it One dimension.} To get an intuition on the scaling function for the mean perimeter and to appreciate the fact 
that its non-monotonicity and singularity at $x=0$ is indeed induced
by the 2D geometry, it is useful to first compute the analogous quantity
in the interesting in itself one dimensional (1D) setting. In 1D, the corresponding quantity 
is the mean span $\langle L^{(d)}_{1\mathrm{D}}(t) \rangle$ of a Brownian motion on a 
semi-infinite line, starting at a distance $d$ from a reflecting origin (see inset of 
Fig.\ref{recap}(c)). It can be expressed as 
\begin{equation}\label{defL1D}
\langle L^{(d)}_{1\mathrm{D}}(t) \rangle=\langle \mathcal{M}^{(d)}_{\mathrm{right}}(t) 
\rangle+\langle \mathcal{M}^{(d)}_{\mathrm{left}}(t) \rangle ,
\end{equation} 
where $ \langle \mathcal{M}^{(d)}_{\mathrm{right}}(t) \rangle$ and $\langle 
\mathcal{M}^{(d)}_{\mathrm{left}}(t) \rangle$ denote the mean maximal extensions of the walk 
up to time $t$ to the right and 
to the left respectively.

For convenience, we shift the reflecting wall to $-d$ and consider the walk starting
at the origin $0$. We compute the right and the left parts separately, starting with
the right. After integration by parts, it is obtained that
$\langle \mathcal{M}^{(d)}_{\mathrm{right}}(t)
\rangle= \int_0^{\infty} \diff{y}\, \left(1- S^{(d)}_{\rm right}(t|y)\right)$, where $S^{(d)}_{\rm right}(t|y)= 
{\rm Prob.}[\mathcal{M}^{(d)}_{\mathrm{right}}(t)\le y]$ is the cumulative distribution of 
$\mathcal{M}^{(d)}_{\mathrm{right}}(t)$, given that the particle starts at the origin. In turn, the
cumulative distribution $S^{(d)}_{\rm right}(t|y)$ is just the probability that the walker
starting at the origin stays within the box $[-d,y]$ up to time $t$, i.e., the survival
probability of the walker with a reflecting wall at $-d$ and an absorbing wall at $y>0$. 
In contrast, for the left side, the maximal displacement can at most be $d$, due to
the presence of the reflecting wall and one 
finds
$\langle \mathcal{M}^{(d)}_{\mathrm{left}}(t)
\rangle= \int_{-d}^{0} \diff{y}\, \left(1- S_{\rm left}(t|y)\right)$, where $S_{\rm left}(t|y)$
is the survival probability of a walker in the semi-infinite region $[y,\infty[$ with
an absorbing wall at $-d<y<0$. These survival probabilities can be computed using standard
techniques~\cite{Redner,Bray:2013}. They are best expressed in terms of Laplace transforms with
respect to $t$. Denoting by $\hat{f}(p)=\int_0^{\infty} f(t)\, e^{-p\,t}\, \diff{t}$ the
Laplace transform of a function $f(t)$, we find
\begin{eqnarray}
&\left\langle \hat{\mathcal{M}}^{(d)}_{\textrm{right}}(p) \right\rangle &= 
\int_0^{+\infty} \mathrm{d}y \left(\frac{1}{p}-\hat{S}^{(d)}_{\textrm{right}}(p|y) \right) \nonumber \\
&\left\langle \hat{\mathcal{M}}^{(d)}_{\textrm{left}}(p) 
\right\rangle &= \int_{-d}^0 \mathrm{d}y \left(\frac{1}{p}-\hat{S}_{\textrm{left}}(p|y)\right) 
\end{eqnarray}
where
\begin{eqnarray}
&&\hat{S}^{(d)}_{\textrm{right}}(p|y)=\frac{1}{p} 
\left( 1-\frac{\cosh\left(\sqrt{\frac{p}{D}}d\right)}{\cosh \left( \sqrt{\frac{p}{D}}(y+d)\right)}\right) 
\nonumber  \\
&&\hat{S}_{\textrm{left}}(p|y)=\frac{1}{p} \left( 1-\mathrm{e}^{\sqrt{\frac{p}{D}}y}  \right).
\end{eqnarray}
Inverting the Laplace transforms, we get
\begin{eqnarray}
&&\kern-1em\tilde{\mathcal{M}}_{\mathrm{right}}(x) \equiv 
\left\langle \frac{\mathcal{M}^{(d)}_{\mathrm{right}}(t)}{\sqrt{Dt}} \right\rangle \nonumber \\
&& = \frac{2}{\sqrt{\pi}} -4\sum\limits_{n=1}^{+\infty} 
\frac{(-1)^n}{4n^2-1} \left( \frac{\e{-n^2x^2}}{\sqrt{\pi}} -nx \erfc{nx} \right) \\
&&\kern-1em\tilde{\mathcal{M}}_{\mathrm{left}}(x) 
\equiv \left\langle \frac{\mathcal{M}^{(d)}_{\mathrm{left}}(t)}{\sqrt{Dt}} 
\right\rangle = \frac{2}{\sqrt{\pi}} \left(1-\e{-\frac{x^2}{4}}\right) + x \erfc{\frac{x}{2}} \nonumber \\
\label{M1D}
\end{eqnarray}
with $x=d/\sqrt{Dt}$ denoting the scaled distance. This gives, from Eq.~\eqref{defL1D},
$
\langle L^{(d)}_{1\mathrm{D}}(t) \rangle= \sqrt{D\,t}\,\, {\tilde L}_{1\mathrm{D}}\left(\frac{d}{\sqrt{D\,t}}\right)
$
where the 1D scaling function, ${\tilde L}_{1\mathrm{D}}(x)=  
\tilde{\mathcal{M}}_{\mathrm{right}}(x)+ \tilde{\mathcal{M}}_{\mathrm{left}}(x)$,
is plotted in Fig.~\ref{recap}(c).
The scaling function increases  
monotonically with $x$ from ${\tilde L}_{1\mathrm{D}}(x\to 0)=\sqrt{\pi}$ (when the walker
starts at the wall) to ${\tilde L}_{1\mathrm{D}}(x\to \infty)=4/\sqrt{\pi}$ (where
the walker does not feel the wall and one recovers the mean span of the walker
in the absence of the wall).

Several conclusions can be drawn from these expressions (see 
Fig.~\ref{recap}(c)): (i) The presence of the reflecting wall preserves
the diffusive scaling of the mean perimeter $\sim \sqrt{D\,t}$.  
(ii) However, for fixed $d$ and late times, $\left\langle L^{(d)}_{\mathrm{1D}}(t)/\sqrt{Dt} 
\right\rangle\to {\tilde L}_{1\mathrm{D}}(x\to 0)=\sqrt{\pi}$, which is lower 
than $4/\sqrt{\pi}$ obtained in the absence of confinement (\mbox{$d\to \infty$}).  (iii) The mean 
perimeter is minimized when the walker starts from the reflecting wall and (iv) 
is an increasing analytic function of the distance from the reflecting 
wall. 

We show below that in 2D, while points (i) and (ii) continue to hold,
(iii) and (iv) are no longer valid. Our 
main results are indeed the non-monotonicity of the mean perimeter of the convex 
hull in 2D and its non-analyticity for small starting distances from the 
reflecting wall. These two features are in striking contrast with the 
one-dimensional case derived above (see Fig.\ref{recap}).

{\em Two dimensions.} We now turn to the 2D case, where we consider a Brownian motion in a semi-infinite 
medium (delimited by a reflecting wall) starting at 
a distance $d$ from this wall (see Fig.~\ref{recap}(a)). 
To compute the mean perimeter of the convex hull of the walk of duration $t$, we follow
the general set-up developed in Refs.~\cite{MajumdarPRL09,Majumdar:2010}. By adapting
Cauchy's formula~\cite{Cauchy:1832} for the perimeter of closed convex curves to random curves,
it was shown that the mean perimeter of the convex hull of any two dimensional stochastic
process, including that of a Brownian motion, can be expressed as~\cite{MajumdarPRL09,Majumdar:2010}
\begin{eqnarray}
\langle L^{(d)}_{2 \mathrm{D}}(t) \rangle =  \int_0^{2 \pi} \diff{\theta} \langle 
\mathcal{M}^{(d)}(\theta,t) \rangle 
\label{defL}
\end{eqnarray} 
where $\mathcal{M}^{(d)}(\theta,t)$ is the maximal projection of the trajectory
of the walker up to time $t$ in 
the direction $\theta$. As in 1D, it is useful to express the mean in terms of the
cumulative distribution of $\mathcal{M}^{(d)}(\theta,t)$ 
\begin{eqnarray}
\langle
\mathcal{M}^{(d)}(\theta,t) \rangle
=\int_0^{+\infty} \kern-0.5em \diff{M} \left(1-S^{(d)}(t|M,\theta)\right)
\label{maxtheta}
\end{eqnarray}
where the cumulative distribution $S^{(d)}(t|M,\theta)$ can be identified to the survival probability up to 
time $t$ of the walker in the semi-infinite plane, but bounded additionally by an absorbing
infinite wall at distance $M$ from the starting point, 
perpendicular to the direction $\theta$ (see inset of Fig.~\ref{recap}(b)). It then defines an infinite wedge 
of top angle $\alpha/2=\pi/2-\theta$ with one absorbing edge and one reflecting edge. However, by
adding a twin wedge symmetrically around the reflecting edge,
the survival probability of the walker in the original wedge is the same as
in the `doubled' wedge
with twice the top angle $\alpha=\pi-2\,\theta$, but with two absorbing edges. The survival
probability of a Brownian walker, starting initially at the polar coordinates $(r_0,\varphi_0)$, inside a wedge of angle 
$\alpha$ with two absorbing edges is~\cite{Redner}  
\begin{eqnarray}\label{survie}
&& S(t|r_0,\varphi_0)=\dfrac{r_0}{\sqrt{\pi Dt}} \;e^{-\frac{r_0^2}{8Dt}} 
\sum\limits_{m=0}^{+\infty} \dfrac{\sin\left(\frac{(2m+1)\pi \varphi_0}{\alpha}\right)}{2m+1} \nonumber \\
&&\quad \times \left[ \mathrm{I}_{\frac{(2m+1)\pi}{2 \alpha}-\frac{1}{2}} 
\left(\frac{r_0^2}{8Dt}\right)+ \mathrm{I}_{\frac{(2m+1)\pi}{2 \alpha}+
\frac{1}{2}} \left(\frac{r_0^2}{8Dt}\right) \right]
\end{eqnarray}
where $\mathrm{I}_{\nu}(z)$ is the standard modified Bessel function. The initial position $(r_0,\varphi_0)$
can be expressed in terms of the original variables $M$, $d$ and $\theta$ (see inset of Fig.\ref{recap}(b)). For convenience,
we introduce two new dimensionless variables
as \mbox{$u\equiv M/\sqrt{Dt} = r_0  \sin\varphi_0/\sqrt{Dt}$} and \mbox{$x\equiv d/\sqrt{Dt} =r_0  \sin(\alpha/2-\varphi_0)/\sqrt{Dt}$} where we recall that \mbox{$\alpha=\pi-2\theta$}. Then it follows that
\begin{eqnarray}
&&\frac{r_0}{\sqrt{Dt}}=\dfrac{1}{\cos\theta} \sqrt{x^2+2 x u \sin\theta+u^2} \label{r0}\\
&&\varphi_0=\arccos\left( \dfrac{x+u \sin\theta}{\sqrt{u^2+2xu\sin\theta+x^2}}\right). \label{phi0}
\end{eqnarray}
Plugging the result for the survival probability in Eq.~\eqref{survie} (with $r_0$ and $\varphi_0$
expressed as functions of $u$ and $x$ via Eqs.~\eqref{r0} and \eqref{phi0}) into Eq.~\eqref{maxtheta}
we get
\begin{eqnarray}
\label{Mthetax}
&&\tilde{\mathcal{M}}(\theta,x) \equiv \left\langle
\frac{\mathcal{M}^{(d)}(\theta,t)}{\sqrt{Dt}} \right\rangle
=\int_0^{+\infty} \mathrm{d}u \; \nonumber \\ 
&  & \Bigg\{ 1 -\frac{\sqrt{x^2+2xu\sin\theta+u^2}}{\sqrt{\pi}\cos\theta}   
\mathrm{e}^{-\frac{x^2+2xu\sin\theta+u^2}{8 \cos^2\theta}}\times  \nonumber \\
& & \sum\limits_{m=0}^{\infty}\dfrac{\sin\left((2m+1)\frac{\pi}{\alpha}
\arccos\left( \frac{x+u \sin\theta}{\sqrt{u^2+2xu\sin\theta+x^2}}\right)\right)}{2m+1}  \times \nonumber \\
&&  \left[\mathrm{I}_{\nu}\kern-0.3em\left(\frac{x^2+2xu\sin\theta+u^2}{8 \cos^2\theta}\right) 
\kern-0.3em +\mathrm{I}_{\nu+1}\kern-0.3em\left(\frac{x^2+2xu\sin\theta+u^2}{8 \cos^2\theta}\right) 
\right] \kern-0.5em \Bigg\}  \nonumber \\
\end{eqnarray}
with \mbox{$\nu=(2m+1)\pi/(2\alpha)-1/2$}.
Integrating over $\theta$ in Eq.~\eqref{defL} then provides our final result
for the mean rescaled perimeter (MRP) 
\begin{eqnarray}\label{Lresc}
\tilde{L}_{2\mathrm{D}}(x) \equiv \left\langle \dfrac{L^{(d)}_{2\mathrm{D}}(t)}{\sqrt{Dt}} \right\rangle = 
\int_0^{2\pi}  \mathrm{d}\theta\, \tilde{\mathcal{M}}(\theta,x)
\end{eqnarray}
with $\tilde{\mathcal{M}}(\theta,x)$ given explicitly in Eq.~\eqref{Mthetax}.
As expected, 
the MRP is a function of only one parameter, the rescaled distance to the wall $x=d/\sqrt{Dt}$.

Interestingly, we show in \cite{SI} that the MRP, given in Eqs.~\eqref{Lresc} and \eqref{Mthetax} for arbitrary $x=d/\sqrt{Dt}$,
simplifies a great deal in the important case $x=0$ of a Brownian motion 
starting from the reflecting wall (or equivalently starting at any fixed distance but for large times) into
\begin{equation}
\tilde{\mathcal{M}}(\theta,0) = 2 \sqrt{\pi} \dfrac{\cos\theta}{\pi-2\theta}.
\end{equation}
As a simple check, we recover, for $\theta=0$ (i.e., in the direction parallel
to the reflecting wall), the result for the non-confined case \mbox{$\tilde{\mathcal{M}}(0,0) =2/\sqrt{\pi}$}. Indeed, the potential reflections 
on the wall do not affect the walk in the parallel direction. For $\theta=\pi/2$, outwards orthogonally 
to the wall, we recover the 1D result obtained above $\tilde{\mathcal{M}}\left(\pi/2,0\right)=\sqrt{\pi}$, 
which is higher than in the non-confined 
case. Indeed, the wall pushes the trajectories further in this direction.
Finally, it is straightforward to obtain the MRP of the convex hull by integrating over the angle $\theta$
\begin{equation}
{\tilde L}_{2\mathrm{D}}(0) = 2 \sqrt{\pi} \, \mathrm{Si}(\pi)=6.56495\dots
\end{equation}
where $\mathrm{Si}(z)=\int_0^z \frac{\sin x}{x}\, \diff{x}$.
Note that the MRP for a walk starting from the reflecting wall still grows as $\sqrt{D\,t}$, but 
the prefactor $6.56495\dots$ is lower than the non-confined value $\sqrt{16\, \pi}=7.08982\dots$.

As $x$ increases from $0$, the scaling function ${\tilde L}_{2\mathrm{D}}(x)$ in \eqref{Lresc}, supplemented
by \eqref{Mthetax}, first decreases, achieves a minimum and then increases again before eventually
saturating to the constant $\sqrt{16\, \pi}$ corresponding to the non-confined case (see Fig.~\ref{recap}(b)
for a plot). This non-monotonic behavior can already be understood by
analyzing the $x\to 0^+$ limit. From Eq.~\eqref{Mthetax}, one can show
that for small $x$ (see \cite{SI} for details)
\begin{eqnarray}\label{Mpetitx}
&& \tilde{\mathcal{M}}(\theta,x)-2\sqrt{\pi}\frac{\cos \theta}{\pi-2\theta}+\sin\theta x \nonumber \\
&&\qquad  =\begin{dcases} \frac{\sqrt{\pi}}{2}\frac{\cos \theta}{\pi-2\theta} x^2 + 
\mathcal{O}(x^3) & \textrm{for $\theta>0$}  \\
 \mathrm{C}(\theta) \; x^{2\nu_0+2} +\mathcal{O}(x^2) & \textrm{for $\theta<0$}  \\
\end{dcases} 
\end{eqnarray}
where $\nu_0=\theta/(\pi-2\theta)$ and the amplitude C$(\theta)$ has a complicated
expression (see \cite{SI}), which is negative for all $\theta<0$.
Interestingly, the leading order correction term in \eqref{Mpetitx} is
non-analytic only for $\theta<0$. 
Using Laplace's method, the integration 
over $\theta$ finally leads to
\begin{eqnarray}\label{Lpetitx}
&\kern-1.5em\tilde{L}(x) -2\sqrt{\pi}\, \mathrm{Si}(\pi)  &\equi{x\ll 1} -
\frac{8 \sqrt{2\pi^3}}{3 \Gamma\left(\frac{3}{4}\right)} \left(\frac{x}{\ln (1/x)}\right)^{\frac{3}{2}} 
\end{eqnarray}
Let us make a few remarks: (i) 
While, for any fixed $\theta$, $\tilde{\mathcal{M}}(\theta,x)$ has
a nonzero linear term in $x$ for small $x$, the linear term disappears when integrated over $\theta$.
(ii) 
Strikingly, the MRP is nonanalytic as $x\to 0$ (contrary to the 1D case) and
(iii) it starts decreasing from a 
value at $x=0$ which is lower than its $x\to \infty$ limit, so it must display a minimum, as confirmed in 
Fig~\ref{recap}(b).

The existence of this surprising minimum can be qualitatively discussed by identifying the temporal 
regimes of a Brownian motion starting at a distance $d$ from the reflecting wall: (i) At short times 
(\textit{i.e.} $x \gg 1$), the walker does not see the reflecting wall and the unconfined value of the 
MRP $4\sqrt{\pi}$ is recovered. (ii) After a time of order $d^2/D$ (\textit{i.e.} $x \simeq 1$), the 
reflecting wall starts impacting the trajectory by progressively reducing the space effectively 
accessible to the Brownian motion and thus decreasing the MRP. (iii) Next, these reflected trajectories 
start contributing to the outwards part of the convex hull (with respect to the plane). This effective repulsion is an 
antagonistic effect of the wall that turns out to increase the MRP. Finally, contrary to the 1D case, 
the MRP displays a complex behavior with $x$, involving a minimum. In addition, this minimum is global, 
implying in particular that the MRP is not minimized when the Brownian motion starts from the wall anymore, but for a nontrivial value of $x \approx 0.3$.
\begin{figure}[h]
\centering
\includegraphics[width=220pt]{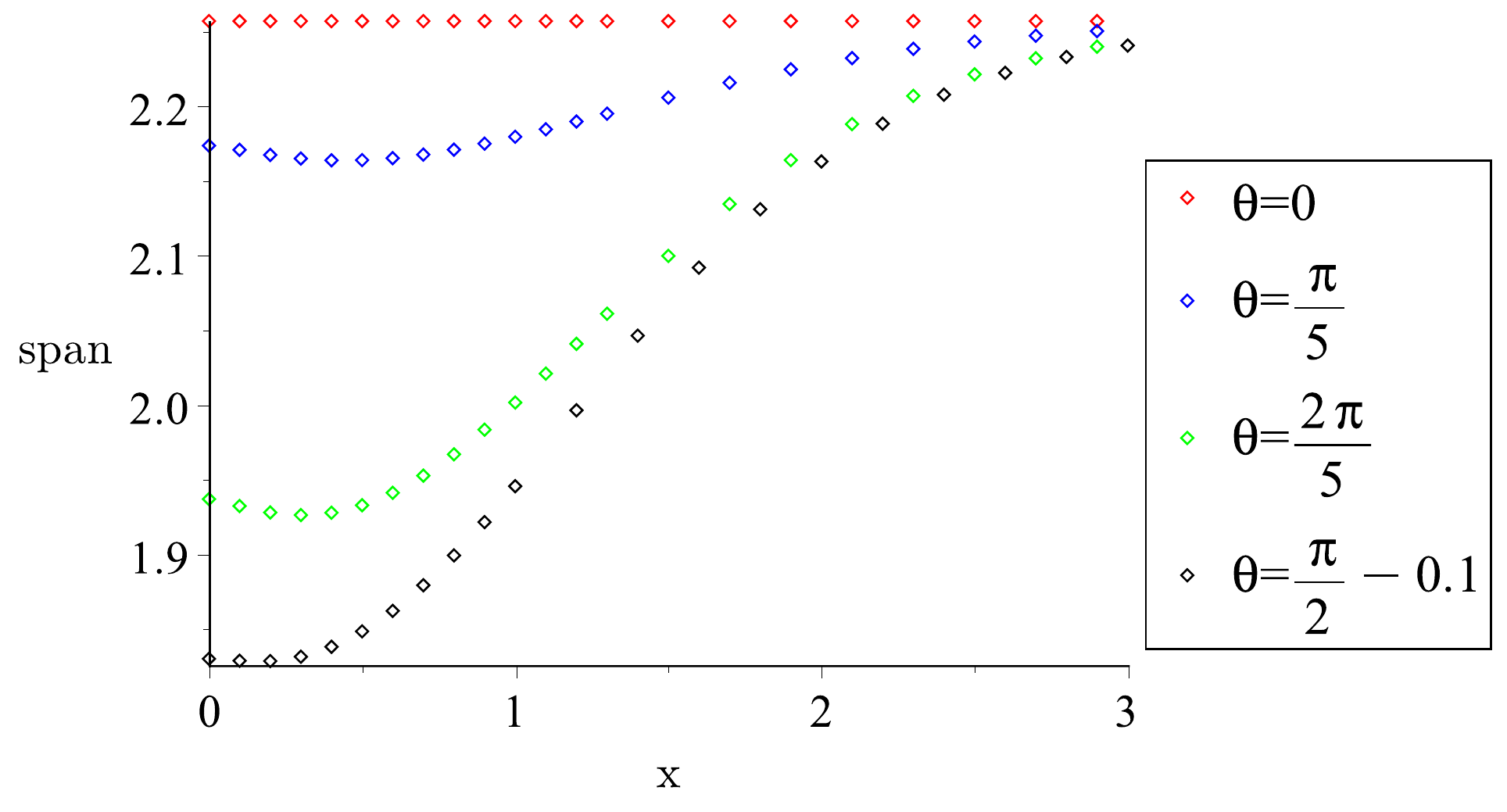}
\caption{Mean 
span in direction $\theta$ as a function of the rescaled distance $x$ for different values of $\theta$. Like the mean perimeter of the convex hull, this quantity 
displays a minimum with respect to $x$ for all directions except parallel or perpendicular to the wall.}
\label{span}
\end{figure}
Insights on this minimum may further be gained by considering the mean span in direction $\theta$ given by 
\mbox{$\langle \mathcal{M}^{(d)}(\theta,t)\rangle+\langle \mathcal{M}^{(d)}(-\theta,t)\rangle$} involved 
in \eqref{defL}, whose small $x$ development is obtained by combining the two forms of \eqref{Mpetitx}. 
Knowing that the mean span in direction $\theta$ at $x=0$ is lower than its large $x$ limit and that it 
starts decreasing ($\mathrm{C}(\theta)<0$), it displays a minimum for $\theta\neq 0$ and $\theta \neq 
\pi/2$ (see Fig. \ref{span}). After integration over $\theta$, the non-monotonicity remains.

In conclusion, we have studied analytically the mean perimeter of the convex hull of a Brownian motion in 1D and 2D in presence of a reflecting wall at the origin. In 2D, this confinement leads to a 
striking non-monotonic and non-analytic (for small distances) behavior of the mean perimeter as a function of the 
scaled starting distance from the wall. The non-monotonicity 2D originates from the competition between two 
antagonistic effects due to the presence of the wall: reduction of the space accessible to the Brownian motion 
and effective repulsion. While these two effects are also at work in 1D, they do not lead to a non-monotonicity. Our work opens up several interesting 
questions for future studies. It would be interesting to know whether this non-monotonicity persists in $d>2$ 
dimensions. Computing the mean area of the convex hull in two dimensions in presence of a reflecting wall remains 
challenging. Finally, it would be interesting to study other forms of confinements, for instance, a Brownian 
motion in an enclosed space or in the presence of an external confining potential.


Support from European Research Council starting Grant No. FPTOpt-277998 is 
acknowledged. SNM acknowledges support by ANR grant
2011-BS04-013-01 WALKMAT.

\section{Derivation of $\tilde{\mathcal{M}}(\theta,0)$}
We start from Eq.~(12) in the main text and first rewrite the $1$ on the right hand side (inside the integral) as
\begin{equation}
\frac{u}{\sqrt{\pi}\cos\theta} \e{-\frac{u^2}{8\cos^2\theta}}
\sum\limits_{m=0}^{+\infty}\frac{(-1)^m}{2m+1} 2
\frac{\e{\frac{u^2}{8\cos^2\theta}}}{\sqrt{2\pi\frac{u^2}{8\cos^2\theta}}}=1.
\end{equation}
A change of variables and the introduction of a regularization parameter $\beta$ yield
\begin{eqnarray}
 && \tilde{\mathcal{M}}(\theta,0) =\dfrac{4}{\sqrt{\pi}} \cos\theta
\sum\limits_{m=0}^{\infty} \dfrac{(-1)^m}{2m+1} \nonumber \\
& & \left. \times \underbrace{\int_0^{+\infty}  \kern-1em \mathrm{d}v \;
\mathrm{e}^{-\beta v} \left( \sqrt{\dfrac{2}{\pi}} \dfrac{\mathrm{e}^v}{\sqrt{v}}-
\mathrm{I}_{\nu}(v)-\mathrm{I}_{\nu+1}(v)\right)}_{A(\beta,m)} \right|_{\beta=1}
\label{regular1}
\end{eqnarray}
where $A(\beta,m)$ can be shown to be given by
\begin{equation}
A(\beta,m)=\sqrt{\dfrac{2}{\beta-1}}-\dfrac{(\beta+\sqrt{\beta^2-1})^{-\nu}}{\sqrt{\beta^2-1}}-
\dfrac{(\beta+\sqrt{\beta^2-1})^{-\nu-1}}{\sqrt{\beta^2-1}}.
\end{equation}

Using next that
\begin{eqnarray}
&\sum\limits_{m=0}^{\infty} \dfrac{(-1)^m a^{-\frac{(2m+1)\pi}{2\alpha}+\frac{1}{2}}}{2m+1} &=\sqrt{a} \arctan\left(a^{-\frac{\pi}{2\alpha}}\right)
\end{eqnarray}
and taking the limit $\beta\to 1$ gives the mean extension in the direction $\theta$
\begin{equation}
\tilde{\mathcal{M}}(\theta,0) = 2 \sqrt{\pi} \dfrac{\cos\theta}{\pi-2\theta}.
\end{equation}

\section{ Asymptotic expansion of $\tilde{\mathcal{M}}(\theta,x)$ for small distances}
We give here the main steps for the derivation of the small $x$ development of $\tilde{\mathcal{M}}(\theta,x)$ stated in Eq.~(16) in the main text. The linear term in $x$ (for \mbox{$-\pi/2<\theta<\pi/2$}) and the quadratic term (for \mbox{$0<\theta<\pi/2$}) are obtained by integration over $u$ of the small $x$ development of the survival probability. In turn, these integrals can be carried out by introducing, as performed in the calculation of the order 0 of $\tilde{\mathcal{M}}(\theta,x)$, a regularization factor $\beta$ (see Eq.~\eqref{regular1}). The integrals involved can be calculated straightforwardly. However, for \mbox{$-\pi/2<\theta<0$}, this method produces a diverging coefficient for the quadratic term, which reveals a non-analytic behavior whose calculation is presented below.

The term $T_0$ responsible for the divergence is
\begin{eqnarray}
&T_0&=\int_0^{+\infty} \kern-1em \diff{u} \left[ \frac{2}{\pi} -\frac{\sqrt{u^2+2xu\sin\theta+x^2}}{\sqrt{\pi}\cos\theta}   \mathrm{e}^{-\frac{u^2+2xu\sin\theta+x^2}{8 \cos^2\theta}} \right. \nonumber \\
& & \qquad \times \cos\left(\frac{\pi}{\alpha}\arccos\left( \frac{u+x \sin\theta}{\sqrt{u^2+2xu\sin\theta+x^2}}\right)\right)  \nonumber \\
&&  \qquad \times \left. \mathrm{I}_{\nu_0}\left(\frac{u^2+2xu\sin\theta +x^2}{8 \cos^2\theta}\right) \right]. \nonumber \\
\end{eqnarray}
with $\nu_0=\theta/(\pi-2\theta)$.
The asymptotic behavior of $T_0$ can be conveniently analyzed by introducing the new variable of integration
\begin{equation}
z=\frac{u^2+2xu\sin\theta+x^2}{x^2\cos^2\theta}.
\end{equation}
It leads to
\begin{eqnarray}
&&\kern-1em T_0=\int_1^{1/\cos^2\theta} \diff{z} \frac{x\cos\theta}{2\sqrt{z-1}} \left[\frac{2}{\pi}-\frac{x}{\sqrt{\pi}} \sqrt{z} \right. \nonumber \\
&& \left. \times  \cos\left(\frac{\pi}{\alpha}\arccos\left(-\frac{\sqrt{z-1}}{\sqrt{z}}\right) \right) \e{-\frac{x^2 z}{8}} \textrm{I}_{\nu_0}\left(\frac{x^2 z}{8}\right) \right] \nonumber \\
&&  +\int_1^{+\infty} \diff{z} \frac{x\cos\theta}{2\sqrt{z-1}} \left[\frac{2}{\pi}-\frac{x}{\sqrt{\pi}} \sqrt{z} \right. \nonumber \\
&& \left. \times  \cos\left(\frac{\pi}{\alpha}\arccos\left(\frac{\sqrt{z-1}}{\sqrt{z}}\right) \right) \e{-\frac{x^2 z}{8}} \textrm{I}_{\nu_0}\left(\frac{x^2 z}{8}\right) \right] \nonumber \\
&&\equiv I_1+I_2.
\end{eqnarray}

The first integral $I_1$ can be written as
\begin{eqnarray}
&&I_1=-\frac{2x}{\pi}\sin\theta -\frac{x^2 \cos\theta}{2\sqrt{\pi}} \int_1^{1/\cos^2\theta} \diff{z} \sqrt{\frac{z}{z-1}} \nonumber \\
&& \qquad \times \cos\left(\frac{\pi}{\alpha}\arccos\left(-\frac{\sqrt{z-1}}{\sqrt{z}}\right) \right) \e{-\frac{x^2 z}{8}} \textrm{I}_{\nu_0}\left(\frac{x^2 z}{8}\right). \nonumber \\
\end{eqnarray}
The second integral $I_2$ cannot be split as it leads to two diverging contributions. Subtracting to the integrand the order 0 given by
\begin{eqnarray}
&&\int_0^{+\infty} \diff{z} \frac{x\cos\theta}{2\sqrt{z}} \left[ \frac{2}{\pi} -\frac{x}{\sqrt{\pi}} \sqrt{z} \e{-\frac{x^2 z}{8}} \textrm{I}_{\nu_0}\left(\frac{x^2 z}{8}\right) \right] \nonumber \\
&& \quad = 4 \, \nu_0 \frac{\cos\theta}{\sqrt{\pi}},
\end{eqnarray}
we obtain
\begin{eqnarray}
 &&\kern-1em I_2=4 \, \nu_0 \frac{\cos\theta}{\sqrt{\pi}}+\int_1^{+\infty} \diff{z} \frac{x\cos\theta}{\pi} \left(\frac{1}{\sqrt{z-1}}-\frac{1}{\sqrt{z}} \right) \nonumber \\
&&\kern-1em -\frac{x^2\cos\theta}{2\sqrt{\pi}} \int_1^{+\infty} \kern-1em \diff{z} \kern-0.5em\left[\sqrt{\frac{z}{z-1}} \cos \kern-0.3em\left(\frac{\pi}{\alpha} \arccos\left(\sqrt{\frac{z-1}{z}}\right) \kern-0.3em\right)-1 \right] \nonumber \\
&& \qquad \times \e{-\frac{x^2 z}{8}} \textrm{I}_{\nu_0}\left(\frac{x^2 z}{8}\right) \nonumber \\
&&\kern-1em -\int_0^1 \diff{z} \frac{x\cos\theta}{2\sqrt{z}} \left[ \frac{2}{\pi} -\frac{x}{\sqrt{\pi}} \sqrt{z} \e{-\frac{x^2 z}{8}} \textrm{I}_{\nu_0}\left(\frac{x^2 z}{8}\right) \right].
\end{eqnarray}
This finally yields 
\begin{eqnarray}\label{T0}
&&T_0=\frac{4\theta\cos\theta}{\sqrt{\pi}(\pi-2\theta)}-\frac{2x}{\pi}\sin\theta +\mathrm{C}(x,\theta)
\end{eqnarray}
with
\begin{eqnarray}\label{defCx}
&&\mathrm{C}(x,\theta)=\frac{x^2\cos\theta}{2\sqrt{\pi}} \left\{\int_0^1 \diff{z} \e{-\frac{x^2 z}{8}} \textrm{I}_{\nu_0}\left(\frac{x^2 z}{8}\right) \right. \nonumber \\
&& - \int_1^{1/\cos^2\theta} \kern-0.5em \diff{z} \sqrt{\frac{z}{z-1}}\cos\left(\frac{\pi}{\alpha}\arccos\left(-\frac{\sqrt{z-1}}{\sqrt{z}}\right) \right) \nonumber \\
&& \qquad \qquad \times \e{-\frac{x^2 z}{8}} \textrm{I}_{\nu_0}\left(\frac{x^2 z}{8}\right) \nonumber \\
&& -\int_1^{+\infty} \kern-0.5em \diff{z} \left[\sqrt{\frac{z}{z-1}} \cos\left(\frac{\pi}{\alpha} \arccos\left(\sqrt{\frac{z-1}{z}}\right) \right)-1 \right] \nonumber \\
&& \left. \qquad \qquad \times \e{-\frac{x^2 z}{8}} \textrm{I}_{\nu_0}\left(\frac{x^2 z}{8}\right) \right\}.
\end{eqnarray}
The constant and linear terms of Eq.~\eqref{T0} contribute to give the constant and linear terms of Eq.~(16) of the main text. As for the last term, we use the development of the integrand of Eq.~\eqref{defCx} for small $x$
\begin{equation}\label{devbessel}
\e{-\frac{x^2 z}{8}} \textrm{I}_{\nu_0}\left(\frac{x^2 z}{8}\right) \equi{x\to0} \frac{x^{2\nu_0}z^{\nu_0}}{16^{\nu_0} \Gamma(1+\nu_0)} +\mathcal{O}(x^{2+2\nu_0})
\end{equation}
to obtain
\begin{equation}
\mathrm{C}(x,\theta) \equi{x\to0}  \mathrm{C}(\theta) x^{2+2\nu_0}+ \mathcal{O}(x^2)
\end{equation}
where
\begin{eqnarray}\label{defC}
&&\kern-1em \mathrm{C}(\theta)=\frac{\cos\theta}{2\sqrt{\pi} 16^{\nu_0} \Gamma(1+\nu_0)} \Bigg\{ \frac{1}{1+\nu_0}  \nonumber \\
&&\kern-1em - \int_1^{1/\cos^2\theta} \kern-0.5em \diff{z} \sqrt{\frac{z}{z-1}}\cos\left(\frac{\pi}{\alpha}\arccos\left(-\frac{\sqrt{z-1}}{\sqrt{z}}\right) \right) z^{\nu_0} \nonumber \\
&& \kern-1em \left. -\int_1^{+\infty} \kern-1em \diff{z} \kern-0.3em\left[\sqrt{\frac{z}{z-1}} \cos\left(\frac{\pi}{\alpha} \arccos\left(\sqrt{\frac{z-1}{z}}\right) \right)-1 \right] z^{\nu_0} \right\}. \nonumber \\
\end{eqnarray}
Replacing $\nu_0$ and $\alpha$ with their values $\theta/\alpha$ and $\pi-2\theta$ finally leads to the small $x$ development of \mbox{$\tilde{M}(\theta,x)$} for \mbox{$-\pi/2<\theta<0$} (see Eq.~(16) of the main text).
\begin{eqnarray}
&&\tilde{M}(x)=\frac{2\sqrt{\pi}\cos\theta}{\pi-2\theta} -\sin\theta \, x + \mathrm{C}(\theta) x^{2+\frac{2\theta}{\pi-2\theta}} +\mathcal{O}({x^2}) \nonumber \\
\end{eqnarray}

\section{Numerical simulations}

We computed the mean perimeter of the convex hull of a Brownian motion via numerical simulations. We constructed Gaussian random walks of $10^5$ steps with a time step $\Delta \tau=10^{-3}$ when the walker is further than a distance $d\simeq 0.2$ from the wall. When the walker approaches the wall, we take an adapted time step quadratic in the distance $d$ to the wall $\Delta \tau = (0.1 \, d+\lambda)^2$ with $\lambda=0.01$. The parameter $\lambda$ should not be taken too small to prevent the computation time from diverging. The convex hull is then constructed using the Graham scan algorithm (see Ref.~23 of the main text) for each Brownian walk, its perimeter calculated and averaged over $10^5$ realisations. Agreement is found with our analytical prediction (see Fig.1(b) of the main text).


\begin{thebibliography}{99}

\bibitem{Berg:1983} H. C. Berg, {\em Random Walks in Biology} (Princeton University Press, New York, 1983);
L. Edelstein-Keshet, {\em Mathematical Models in biology} (McGraw Hill, Boston, 1988).

\bibitem{Bartumeus:2005} F. Bartumeus et. al., Ecology, {\bf 86}, 3078 (2005).

\bibitem{Murphy:1992} D. Murphy and B. Noon, Ecol. Appl. {\bf 2}, 3 (1992).

\bibitem{Worton:1995}  B. J. Worton, Biometrics {\bf 51}, 1206 (1995).

\bibitem{Giuggioli:2011} L. Giuggioli, J. R. Potts, and S. Harris,
PLoS Comput. Biol. {\bf 7}, e1002008 (2011).

\bibitem{Takacs:1980} L. Tak\'{a}cs, Amer. Math. Month. {\bf 87},
142 (1980).

\bibitem{ElBachir:1983} M. El Bachir, {\it
L'enveloppe convexe du mouvement brownien} (Ph.D thesis, Universit\'e Paul
Sabatier, Toulouse, France, 1983).

\bibitem{Letac:1993} G. Letac, J. Theor. Prob. {\bf 6}, 385 (1993).

\bibitem{MajumdarPRL09}  J. Randon-Furling, S. N. Majumdar, and
A. Comtet, Phys. Rev. Lett. {\bf 103}, 140602 (2009).

\bibitem{Majumdar:2010} S. N. Majumdar, A. Comtet, and
J. Randon-Furling, J. Stat. Phys. {\bf 138}, 955 (2010).

\bibitem{Reymbaut:2011} A. Reymbaut, S. N. Majumdar, and A. Rosso,
J. Phys. A: Math. Theor. {\bf 44}, 415001 (2011).

\bibitem{Dumonteil:2013} E. Dumonteil, S. N. Majumdar, A. Rosso,
and A. Zoia, Proc. Nat. Acad. Sci. USA, {\bf 110}, 4239 (2013).

\bibitem{Randon:2013} J. Randon-Furling, J. Phys. A: Math. Theor. {\bf 46}, 015004 (2013).

\bibitem{Lukovic:2013} M. Lukovi\'c, T. Geisel, and S. Eule,
New. J. Phys. {\bf 15}, 063034 (2013).

\bibitem{Randon:2014} J. Randon-Furling, Phys. Rev. E {\bf 89}, 052112 (2014).

\bibitem{Biane:2011} P. Biane and G. Letac, Journal of Theor. Prob. {\bf 24}, 330 (2011).

\bibitem{Eldan:2011} R. Eldan, arXiv:1211.2443

\bibitem{Kampf:2012} J. Kampf, G. Last, and I. Molchanov, Proc. Amer. Math. Soc. {\bf 140}, 2527 (2012).

\bibitem{Kabluchko:2014} Z. Kabluchko, and D. Zaporozhets, arXiv:1404.6113

\bibitem{Redner} S. Redner,  {\em A guide to first-passage processes} (Cambridge University Press, Cambridge, 2001).

\bibitem{Bray:2013} A. J. Bray, S. N. Majumdar and G. Schehr, Advances in Physics {\bf 62}, 3 (2013).

\bibitem{Cauchy:1832} A. Cauchy, {\it
La rectification des courbes} (M\'{e}moire de l'Acad\'{e}mie des
Sciences, Paris, 1832).

\bibitem{Graham} R. L. Graham, Inf. Porc. Lett. {\bf 1}, 132 (1972).

\bibitem{SI} see Supplementary Materials, M. Chupeau, O. B\'enichou, and S.~N. Majumdar.





\clearpage

\begin{center}

\Large{\textbf{Supplementary Material for ``Convex hull of Brownian motion in confinement"}}

\end{center}
\end{thebibliography}

\end{document}